\documentclass[12pt,tightenlines,showpacs,showkeys]{revtex4-1}
\usepackage{graphicx,color,amssymb}

\begin{document}

\title{Statistical Analysis of the Correlation
between Active Galactic Nuclei and Ultra-High Energy Cosmic Rays}
\author{Hang Bae Kim$^{1,2}$}\email{hbkim@hanyang.ac.kr}
\author{Jihyun Kim$^1$}\email{jihyunkim@hanyang.ac.kr}
\affiliation{$^1$Department of Physics and The Research Institute of
Natural Science, Hanyang University, Seoul 133-791, Korea}
\address{$^2$Korean Institute of Advanced Study (KIAS), Seoul 130-722, Korea}

\begin{abstract}
We develop the statistical methods for comparing two sets
of arrival directions of cosmic rays
in which the two-dimensional distribution of arrival directions is reduced
to the one-dimensional distributions so that the standard one-dimensional
Kolmogorov-Smirnov test can be applied.
Then we apply them to the analysis of correlation
between the ultra-high energy cosmic rays (UHECR)
with energies above $5.7\times10^{19}\;{\rm eV}$,
observed by Pierre Auger Observatory (PAO) and Akeno Giant Air Shower Array (AGASA),
and the active galactic nuclei (AGN) within the distance $100\;{\rm Mpc}$.
For statistical test,
we set up the simple AGN model for UHECR sources
in which a certain fraction of observed UHECR are
originated from AGN within a chosen distance, assuming that
all AGN have equal UHECR luminosity and smearing angle,
and the remaining fraction are from the isotropic background contribution.
For the PAO data, our methods exclude not only a hypothesis
that the observed UHECR are simply isotropically distributed
but also a hypothesis that they are completely originated from the selected AGN.
But, the addition of appropriate amount of isotropic component
either through the background contribution
or through the large smearing effect improves the correlation greatly
and makes the AGN hypothesis for UHECR sources a viable one.
We also point out that restricting AGN within the distance bin of
$40\!-\!60\;{\rm Mpc}$ happens to yield a good correlation
without appreciable isotropic component and large smearing effect.
For the AGASA data, we don't find any significant correlation with AGN.
\end{abstract}

\maketitle

\section{Introduction}

The sources of ultra-high energy cosmic rays (UHECR) still remain in veil,
but the recent observation of the suppression in the energy spectrum of UHECR
above the so-called Greisen-Zatsepin-Kuzmin (GZK) cutoff
($E_{\rm GZK}\sim4\times10^{19}\;{\rm GeV}$)
\cite{Abraham:2008ru,Abbasi:2007sv}
indicates that at least the UHECR with energies above the GZK cutoff
come from relatively close (within the GZK radius $\sim100\;{\rm Mpc}$)
extragalactic sources.
The magnitude and extension of the intergalactic magnetic fields are supposed
not to be large enough to significantly alter the arrival directions of UHECR
with these energies.
Thus, the distribution of arrival directions is correlated with the distribution
of the sources and the arrival direction analysis for the UHECR with energies
above the GZK cutoff can be very useful in identifying their sources.
An important step toward this direction was initiated by
the correlation between arrival directions
of UHECR and nearby active galactic nuclei (AGN)
reported by the Pierre Auger Observatory (PAO) \cite{Cronin:2007zz}.
Though further analysis with more data weakened the significance
of the correlation \cite{Abraham:2007si,:2010zzj},
it still remains as an important issue.
Another important clue for the UHECR sources may come from the study of
the correlation between the UHECR arrival directions and the large scale structures
manifested in the galaxy distribution or the matter distribution.
It was studied by several groups
\cite{:2010zzj,Takami:2009bv,Cuesta:2009ap,Koers:2008ba,Takami:2008ri,
Kashti:2008bw}
and the results are not quite conclusive yet.
The positive result will provide the basis for the further study of correlations
between the UHECR and specific classes of astrophysical objects.

For the study of correlation between the arrival directions of UHECR and
some astrophysical objects, we have to rely on the statistical methods.
This is because the current limit of angular resolution in CR experiments
and a poor understanding of the intergalactic magnetic fields
make it difficult to pin down individual sources of UHECR from their arrival directions.
This also seems to be unavoidable as we consider that currently
the number of astrophysical objects which are candidates for UHECR sources
is very large while the number of observed UHECR events is small yet.

Statistical studies of correlation between the arrival directions of UHECR
and the astrophysical objects was done in many ways
\cite{Cronin:2007zz,Abraham:2007si,:2010zzj,
Nemmen:2010bp,Jiang:2010yc,Mirabal:2010yi,Abbasi:2008md,
Abbasi:2005qy,Gorbunov:2004bs,Singh:2003xr,Smialkowski:2002by,Torres:2002bb}.
In this paper, first we improve the previously used methods and combine them
to obtain more reliable estimates of the significance of correlation.
Our improvement is twofold.
Firstly, while the previous analyses compared the {\em representative values} of
the two-dimensional distributions of UHECR arrival directions,
for example the number of UHECR events within the certain angular distance
from the supposed sources, ours compares the {\em one-dimensional distributions}
obtained from the two-dimensional distributions,
which conveys more information in the sense that,
compared with the above example of representative value,
this one-dimensional distribution contains information about
the number of UHECR events within any angular distance,
not just within a certain specific angular distance.
Secondly, the previous analyses focused on the deviation from isotropy
rather than the direct correlation with the supposed sources
by taking as their null hypothesis that the UHECR arrival directions are
isotropically distributed.
Here, we take as a null hypothesis the explicit source model for UHECR
including the isotropic component and focus on the direct correlation
of UHECR with the supposed sources.
The basic idea is that we reduce the two-dimensional distribution of arrival
directions to the one-dimensional distributions, which can be compared
by using the well-known Kolmogorov-Smirnov (KS) test.
Then, we apply these methods to examine the correlation between UHECR
arrival directions and AGN.
For statistical test,
we set up the simple AGN model for the UHECR sources, in which
all or a fraction of UHECR above a certain energy cutoff are originated
from the AGN within a certain distance cut.
For simplicity, we assume that all selected AGN
have the equal luminosity and smearing angle of UHECR.
The remaining fraction is filled with the isotropic component of UHECR
to account for the contribution from the outside of the distance cut.
We use the AGN data listed in the 12th edition of
V\'eron-Cetty and V\'eron catalog \cite{VCV}
and the UHECR data obtained by PAO \cite{Abraham:2007si} and
Akeno Giant Air Shower Array (AGASA) \cite{Hayashida:2000zr}
to cover both the southern
and northern hemispheres and compare the results drawn from them.

This paper is organized as follows.
In section \ref{sec2}, we introduce three statistical methods for comparing
two distributions of arrival directions.
In section \ref{sec3}, we explain the simple AGN model for the UHECR sources
and the details needed for the generation of Monte-Carlo events for the model
and the statistical comparison with the observed data.
Section \ref{sec4} presents the results of our analysis.
We conclude in section \ref{sec5}.

\section{Statistical Comparison of Two Arrival Direction Distributions}
\label{sec2}

We want to measure the correlation between the observed UHECR arrival directions
and the directions of the candidate point sources
to examine how plausible the candidate sources are.
What we actually obtain through statistical analysis is the probability that
the observed UHECR arrival direction distribution come from
the expected UHECR arrival direction distribution from a given hypothesis
about the sources.
This comes from statistical comparison of two arrival direction distributions.
Quantifying how similar two distributions of objects on the sphere are
is a non-trivial problem.
Various methods have been discussed
\cite{Takami:2009bv,Takami:2008ri,Koers:2008ba,Kashti:2008bw,Harari:2008zp,
Gorbunov:2007ja,Kalashev:2007ph,Cuoco:2005yd,Kachelriess:2005uf,Sigl:2004yk,
Singh:2003xr,Smialkowski:2002by,Sommers:2000us,Evans:2001rv,Waxman:1996hp}.
Because the distribution on the sphere is two dimensional,
we may apply the two-dimensional version of KS test
\cite{KS2D1,KS2D2,Harari:2008zp}.
The drawback of naive two-dimensional KS test is that
it does not properly care the rotational symmetry
that the distribution on the sphere has
\cite{Lopes:2007zz,Metchev:2002ar,SASD}.
In this paper, we adopt the strategy that we reduce the two-dimensional distribution
on the sphere to one-dimensional one for which the rotational
symmetry is kept manifestly and apply the one-dimensional KS test
to quantify the probability.
For the reduction of the distribution on the sphere to one-dimensional distribution,
we consider three ways described below.

\subsection{Correlational Angular Distance Distribution}

This is most useful when we consider the set of point sources for UHECR.
By the correlation between two different sets of objects,
we mean that nearby each object of one set,
we can find objects of the other set at a certain level.
To quantify the distance on the sphere, we use the angular distance $\theta$ between
two objects on the sphere which is the angle between two unit vectors
$\hat{\bf r}$ and $\hat{\bf r}'$ pointing two objects,
that is, $\cos\theta=\hat{\bf r}\cdot\hat{\bf r}'$.
For each object of the reference set, say the point source set,
we count the number of objects of the other set, say the UHECR arrival direction set,
within a certain angular distance $\theta$.
Then we sum up counts over all objects of the reference set
and estimate the significance level of this number.
This way of counting has been used previously in many literature
\cite{Cronin:2007zz,Abraham:2007si,Gorbunov:2007ja,Gorbunov:2004bs}.

To quantify the significance level of the correlation,
we need to know the probability distribution of the expected number of objects.
It can be calculated analytically or obtained by Monte-Carlo simulation,
once a hypothesis about the UHECR sources is set up.
The simple test bed is a hypothesis that the UHECR arrival directions are
isotropically distributed.
Then the correlation between the set of sources
and the set of UHECR arrival directions was claimed based on
that we observe more count of UHECR than that expected in the simple isotropic
distribution.
But what this procedure actually proves is that the observed UHECR
arrival directions are not isotropically distributed.
To make a stronger claim than simply something more than isotropy
we should use an elaborated hypothesis for the test.

\begin{figure}
\setlength{\unitlength}{1mm}
\thicklines
\begin{picture}(160,80)(0,-6)
\put(28,-7){(a)}
\put(108,-7){(b)}
\put(0,0){\includegraphics[width=60mm]{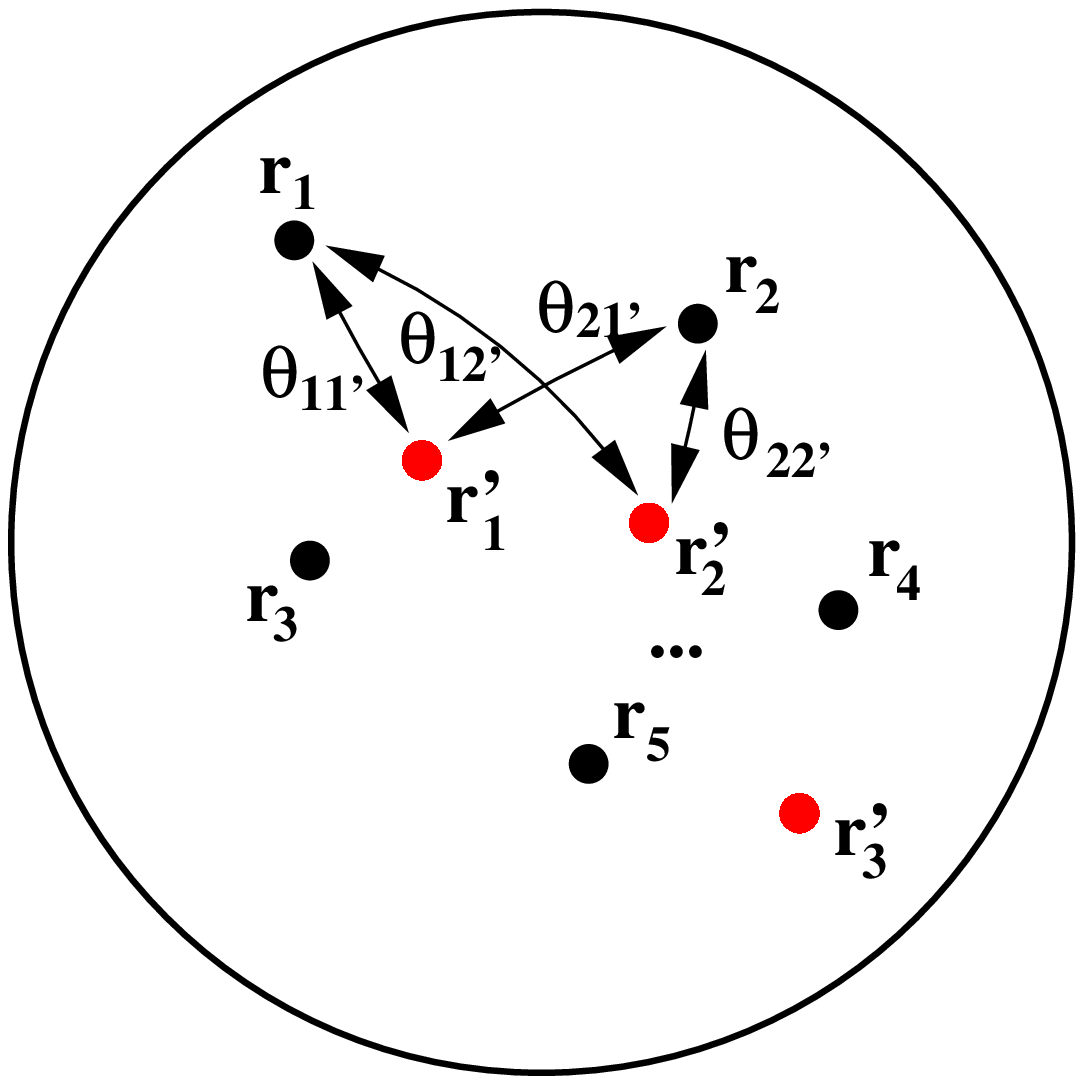}}
\put(65,0){\includegraphics[width=40mm]{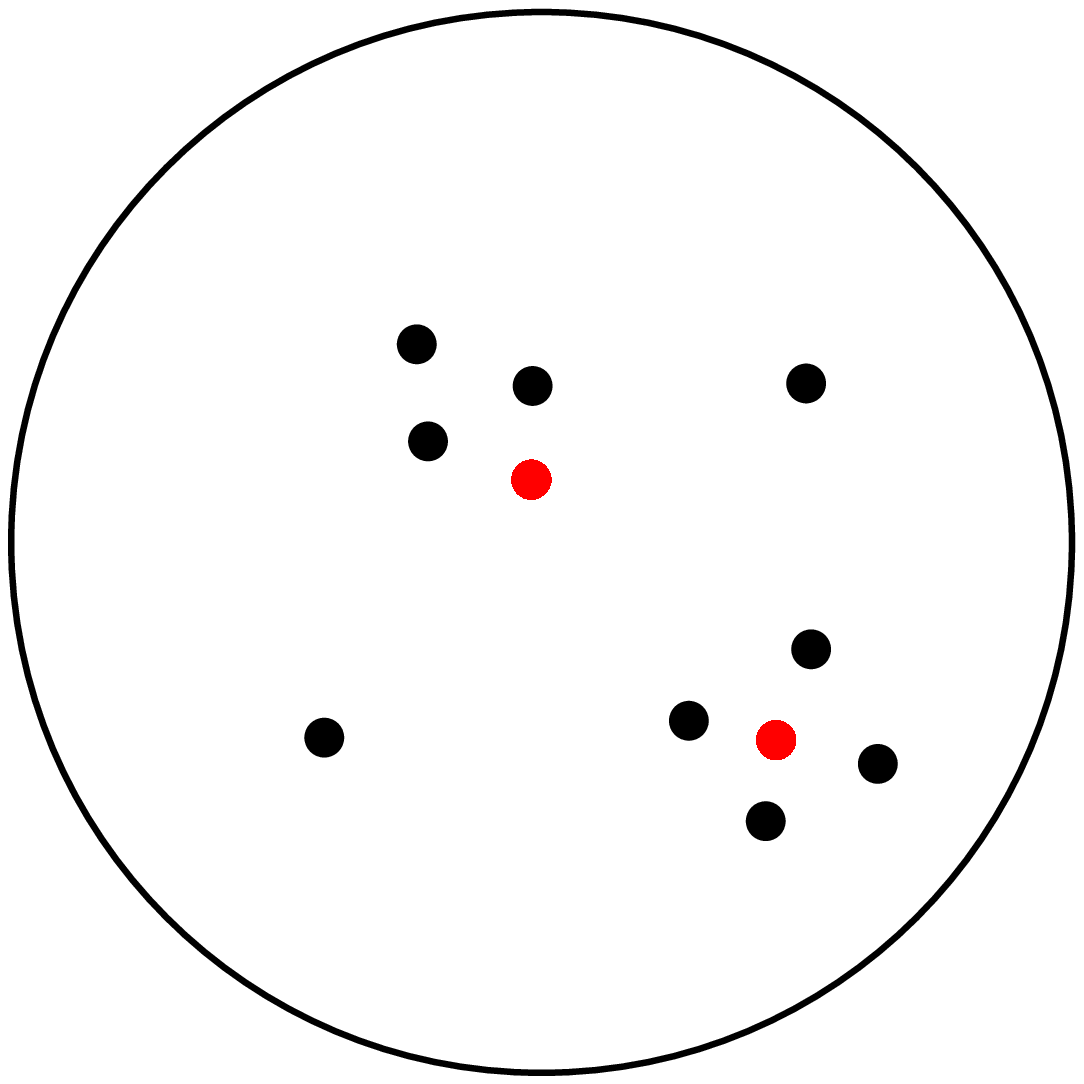}}
\put(90,30){\includegraphics[width=40mm]{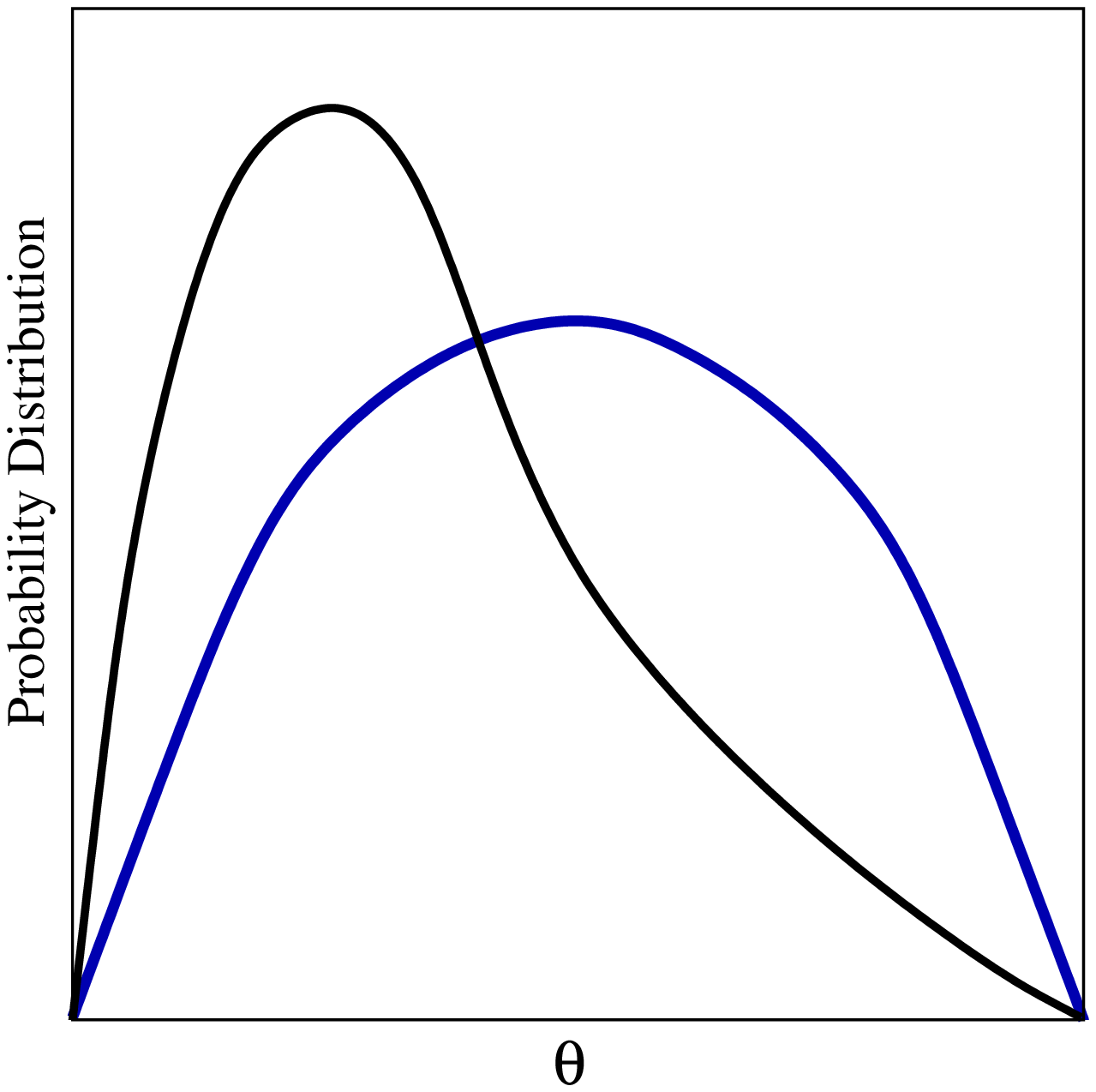}}
\put(115,0){\includegraphics[width=40mm]{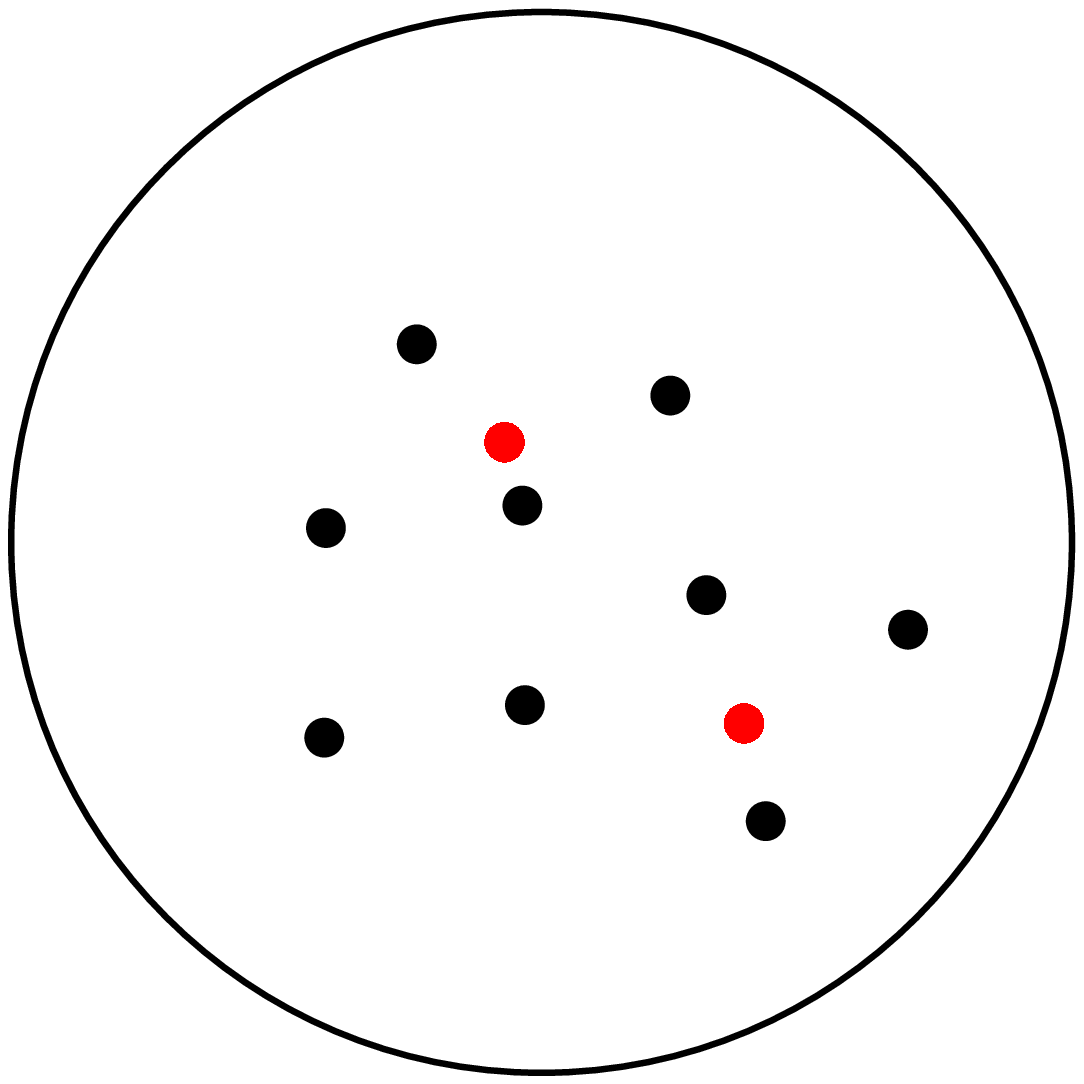}}
\put(71,30){{\bf Correlated}}
\put(131,30){{\bf Isotropic}}
\put(75,33){\vector(1,1){20}}
\put(140,33){\vector(-1,1){16}}
\end{picture}
\caption{Illustrations showing the basic idea of comparing CADD.
(a) CADD is the distribution of all angular distances between the reference
directions (red dots) and the UHECR arrival directions (black dots).
(b) When the observed UHECR events are more clustered around the reference directions
than, say, those of the isotropic distribution, the observed CADD has more counts
at small angles than that expected from the isotropic distribution.}
\label{cadd}
\end{figure}

The above idea can be slightly generalized to estimate the significance level
of the correlation between two sets of objects on the sphere.
Instead of fixing the angular distance $\theta$ to a certain value,
we consider the distribution of the count over the whole range
of the angular distance, that is, from $0^\circ$ to $180^\circ$.
We call it the correlational angular distance distribution (CADD):
\begin{equation}
\hbox{CADD : }
\left\{ \cos\theta_{ij'}\equiv \hat{\bf r}_i\cdot\hat{\bf r}'_j
\;|\; i=1,\dots,N;\; j=1,\dots,M \right\},
\end{equation}
where $\hat{\bf r}_i$ are the UHECR arrival directions,
$\hat{\bf r}'_j$ are the reference directions,
and $N$ and $M$ are total numbers of them, respectively.
In FIG.~\ref{cadd}, we illustrate the basic idea and the usage of the CADD.
We compare the CADD obtained from the reference set (presumably the set
of point sources) and the observed UHECR arrival directions
with that obtained from the same reference set and
the simulated UHECR arrival directions from a given hypothesis.
Because two distributions to be compared are one-dimensional,
the KS test can be applied to yield the probability that the observed
distribution of arrival directions comes from the hypothesis.
The advantage of this method is that it is very simple in practice and
does not require arbitrary angular binning.
Listing in ascending order of all angular distances between pairs of
objects in the reference set and the UHECR arrival direction set
directly gives the cumulative
probability distribution of the count number which is used in the KS test.
Since we directly deal with the cumulative probability distribution,
binning in angular distance is unnecessary and
the expected cumulative probability distribution from the hypothesis
can be made accurate simply by increasing the number of simulated events enough.

\subsection{Flux Exposure Value Distribution}

Koers and Tinyakov proposed another method
to test the correlation between the sources and the observed UHECR
\cite{Koers:2008ba}.
In this method, the two-dimensional distribution of UHECR arrival directions
is reduced to the one-dimensional distribution of expected flux values
at UHECR arrival directions.
At a given arrival direction, the expected flux value is the product of
the UHECR flux expected from a given hypothesis for the sources
and the exposure function at that direction.
Thereby, we call it the flux exposure value distribution (FEVD):
\begin{equation}
\hbox{FEVD : }
\left\{ F_i\equiv f(\hat{\bf r}_i)h(\hat{\bf r}_i)
\;|\; i=1,\dots,N \right\},
\end{equation}
where $\hat{\bf r}_i$ are the UHECR arrival directions,
$N$ is the total numbers of UHECR,
$f(\hat{\bf r}_i)$ and $h(\hat{\bf r}_i)$ are the UHECR flux
and the exposure function, respectively.
It is also very simple in practice and does not require arbitrary angular binning.
One additional merit of this method is that
it can be used for the continuous source distribution.
For the detailed motivations and merits of this method, see Ref.~\cite{Koers:2008ba}.
Here, we take this method for checking and comparison.

\subsection{Auto-Angular Distance Distribution}

The CADD  and the FEVD with the one-dimensional KS test
respects the rotational symmetry.
But they reduce a two-dimensional distribution on the sphere to one-dimensional
distribution of angular distances and flux values.
It means that the loss of information occurs.
The one-dimensional KS test on the CADD catches the correlation of the
source directions and the CR directions, but it can be faked by, say,
the aggregation of UHECR themselves around a few point sources.
To compensate for this, we may consider the distribution of the angular distances
of all pairs of UHECR arrival directions, which we call the auto-angular distance
distribution (AADD):
\begin{equation}
\hbox{AADD : }
\left\{ \cos\theta_{ij}\equiv \hat{\bf r}_i\cdot\hat{\bf r}_j
\;|\; i,j=1,\dots,N \right\},
\end{equation}
where $\hat{\bf r}_i$ are the UHECR arrival directions
and $N$ is the total numbers of UHECR.
It catches the aggregation or spreading of UHECR arrival directions.
It was used previously to examine the clustering of CR arrival directions
\cite{Abbasi:2004ib,Finley:2003ur,Tinyakov:2001ic,Takeda:1999sg}.
The AADD method may not be so relevant for the correlation analysis
because it does not measure the correlation directly.
However, it is still useful to compare two distributions of arrival directions,
and to test the hypothesis for the sources.
It is quite useful when the number of sources is small or a few strong sources
dominate over the others.

\section{Modeling the Arrival Directions of UHECR}
\label{sec3}

For a statistical test, we need to know the expected distribution
of UHECR arrival directions.
To get the expected distribution from a given model for the UHECR sources,
Monte-Carlo simulation is commonly used.
In this section, we present some details needed for the generation
of UHECR arrival directions by using Monte-Carlo simulation
and statistical comparison with the observed data.

\subsection{Arrival directions of UHECR}

\begin{figure}
\centerline{\includegraphics[width=100mm]{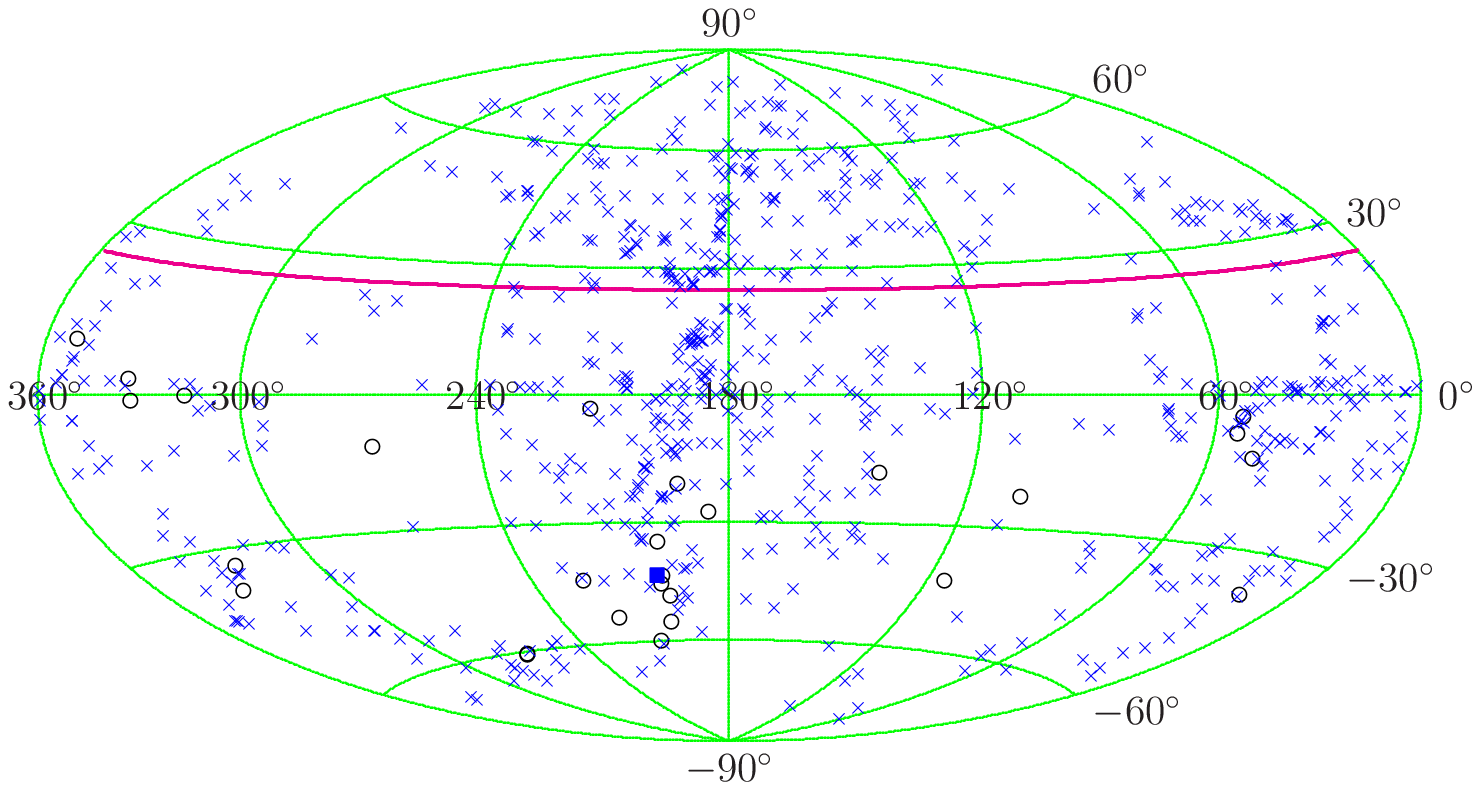}}
\par\vspace{5mm}\par
\centerline{\includegraphics[width=100mm]{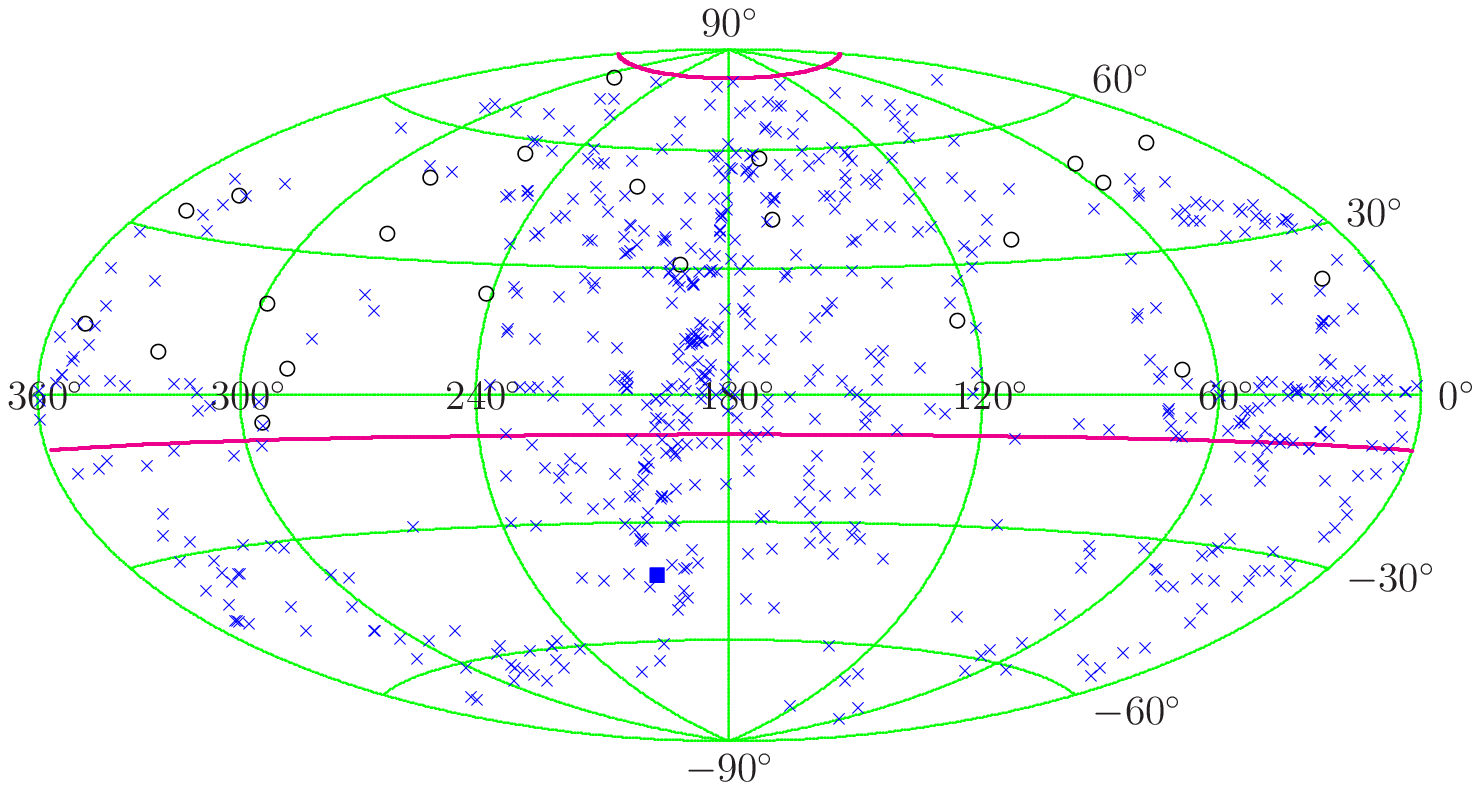}}
\caption{Skymaps of the arrival directions of UHECR,
represented by black circles ({\color{black}$\circ$}),
with energy $E\geq5.7\times10^{19}\,{\rm eV}$
observed by PAO (upper panel) and AGASA (lower panel)
in the equatorial coordinates plotted using the Hammer projection.
The solid red lines are the boundaries of the sky covered by
PAO and AGASA experiments.
The blue crosses ({\color{blue}$\times$}) represent the locations of
AGN with distance $d \leq 100\,{\rm Mpc}$ from the VCV catalog.
The blue square ({\color{blue}$\blacksquare$}) shows the location of Centaurus A.}
\label{skymap-all}
\end{figure}

For statistical analysis of arrival directions,
we use the UHECR data reported
by PAO \cite{Abraham:2007si} and
by AGASA \cite{Hayashida:2000zr}.
These data sets were selected with certain energy cuts.
The energy cuts of the reported data sets are
$5.7\times10^{19}\,{\rm eV}$ for the PAO data set and
$4.0\times10^{19}\,{\rm eV}$ for the AGASA data set, respectively.
The merits of taking a higher value of energy cut for UHECR data
in the analysis of positional correlations with astrophysical objects are
1) we can minimize the deflection due to the galactic and extragalactic
magnetic fields;
2) we can reduce the diffuse isotropic components.
The diffuse isotropic components may come from the contributions of our galaxy
and far distant astrophysical object. We can reduce these contribution
by taking the energy cut above the GZK cutoff, restricting possible sources
within the GZK radius which is around $100\ {\rm Mpc}$.
Of course, the drawback is that we have a smaller number of data,
which reduces the statistical power. So we need to compromise.
Here, we apply the energy cut $E \geq E_c=5.7\times10^{19}\,{\rm eV}$
for both data sets.
With this energy cut,
we have 27 UHECR events from PAO and 23 UHECR events from AGASA.
These are shown in FIG.~\ref{skymap-all}.

The detector array does not cover the sky uniformly and we must consider
its efficiency as a function of the arrival direction.
It depends on the location and the characteristics of the detector array.
Here we consider only the geometric efficiency which is determined
by the location and the zenith angle cut of the detector array.
Then the exposure function $h$ depends only on the declination $\delta$,
\begin{equation}
h(\delta) = \frac{1}{\pi}\left[ \sin\alpha_m\cos\lambda\cos\delta
	+\alpha_m\sin\lambda\sin\delta\right],
\end{equation}
where $\lambda$ is the latitude of the detector array,
$\theta_m$ is the zenith angle cut, and
\[
\alpha_m=\left\{\begin{array}{ll}
0,            & \hbox{for\ } \xi > 1, \\
\pi,          & \hbox{for\ } \xi < -1, \\
\cos^{-1}\xi, & \hbox{otherwise}
\end{array}\right.
\ \hbox{with}\
\xi=\frac{\cos\theta_m-\sin\lambda\sin\delta}{\cos\lambda\cos\delta}.
\]
The PAO site has $\lambda=-35.20^\circ$
and the zenith angle cut of the data is $\theta_m=60^\circ$.
The AGASA site has $\lambda=+35.78^\circ$
and the zenith angle cut of the data is $\theta_m=45^\circ$.

\subsection{The Simple AGN Model for UHECR Sources}

Now we consider a hypothesis
that all or a certain fraction of UHECR with energy above a certain energy cut $E_c$
are originated from the AGN within a certain distance cut $d_c$.

For the correlation analysis, we use AGN
listed in the V\'eron-Cetty and V\'eron (VCV) catalog.
The 12th edition of VCV catalog contains 108,080 quasars ($M<-23$),
BL Lacertae objects and AGN ($M>-23$)
\cite{VCV}.
The catalog includes a variety of contents of each object, for example,
the equatorial coordinates, magnitude, color index, redshift factor and so on.
The catalog may not be a complete list of AGN we need.
But we believe it covers enough number of AGN
so that the statistical analysis will yield the useful information.

Because we apply the energy cut to UHECR data which is higher than the GZK cutoff,
most of probable sources of them are expected to lie within the GZK radius.
With this fact in mind and for the comparison
with Refs.~\cite{Cronin:2007zz,Abraham:2007si},
we pick up the AGN with the distance less than or equal to 100 Mpc
(corresponding to the redshift $z \leq 0.024$) for the correlation test.
The original number of AGN within 100 Mpc in the VCV catalog is 694.
This includes 7 AGN with zero redshift,
which are problematic to be included in our analysis.
Thus, we eliminate these seven AGN from our AGN data set
and the remaining 687 AGN will be used in our analysis.
We show their locations in FIG.~\ref{skymap-all}.

The UHECR luminosity would vary AGN by AGN in general.
We now have a unified picture of AGN
as the supermassive black hole surrounded by accreting materials.
It is also known that the efficient particle acceleration can take place
in the vicinity of the black hole and in the jet up to energies
above $10^{19}\;{\rm eV}$ through several different mechanisms
\cite{Rieger:2009kf}.
However, it is still difficult to infer or to model
the expected UHECR fluxes of all AGN
due to limited measurements done so far and theoretical uncertainties.
Thus, without a priori information,
we assume that all AGN have the equal UHECR luminosity.
Then, the UHECR flux from each AGN is just inversely proportional to
the square of the distance to it.

We are considering AGN as the point sources of UHECR.
However, the trajectories of UHECR are bent by intervening magnetic fields.
Thus, the arrival directions of UHECR deviate from the exact AGN positions.
(We also need to consider the uncertainty coming from the limitation of
the angular resolutions of detector arrays.)
To take this effect into account, we consider AGN as
the smeared sources of UHECR.
The smearing effect also varies AGN by AGN as well as UHECR by UHECR.
But, for simplicity, we assume that all AGN are smeared sources
having gaussian flux distribution about the locations
with a common angular width $\theta_s$, called the smearing angle.
Here, the smearing angle, $\theta_s$, is taken to be a free parameter.
Semi-analytic analysis and numerical simulations of large scale structure
formation indicated that the deflection of UHECR with
$E\ge4.0\times10^{19}\;{\rm eV}$
is at best $2^\circ\sim3^\circ$ \cite{Kotera:2008ae,Dolag:2004kp}.
However, we take $6^\circ$ as a fiducial value of $\theta_s$
to get the rather conservative constraint on the simple AGN model \cite{Kashti:2008bw}.

Now the UHECR flux distribution of each AGN is given by
\begin{equation}
f_i(\theta)=\frac{L}{4\pi d_i^2}\cdot\frac{e^{-(\theta/\theta_s)^2}}{\pi\theta_s^2}\;,
\end{equation}
where $L$ is the common UHECR luminosity,
$d_i$ is the distance to the AGN,
and $\theta$ is the angular distance from the AGN position.
The value of $L$ is related to the total flux of UHECR,
which we do not address here.
Its precise value is not important for our purpose
because it is equal for all AGN and our test compares
the {\em normalized\/} probability distributions of UHECR events over the sphere.

The UHECR with energy above the energy cut $E_c=5.7\times10^{19}\,{\rm eV}$
still can come from the sources lying outside the distance cut $d_c=100\,{\rm Mpc}$.
The fraction of UHECR with $E\ge E_c$ coming from the sources with $d\le d_c$
can be estimated as a function of $E_c$ and $d_c$
by solving the cosmic ray propagation equation.
We denote this fraction as $f_0$.
For $E_c=5.7\times10^{19}\,{\rm eV}$ and $d_c=100\,{\rm Mpc}$,
the estimated value is $f_0\approx0.7$ \cite{Koers:2008ba}.

\begin{figure}
\centerline{\includegraphics[width=100mm]{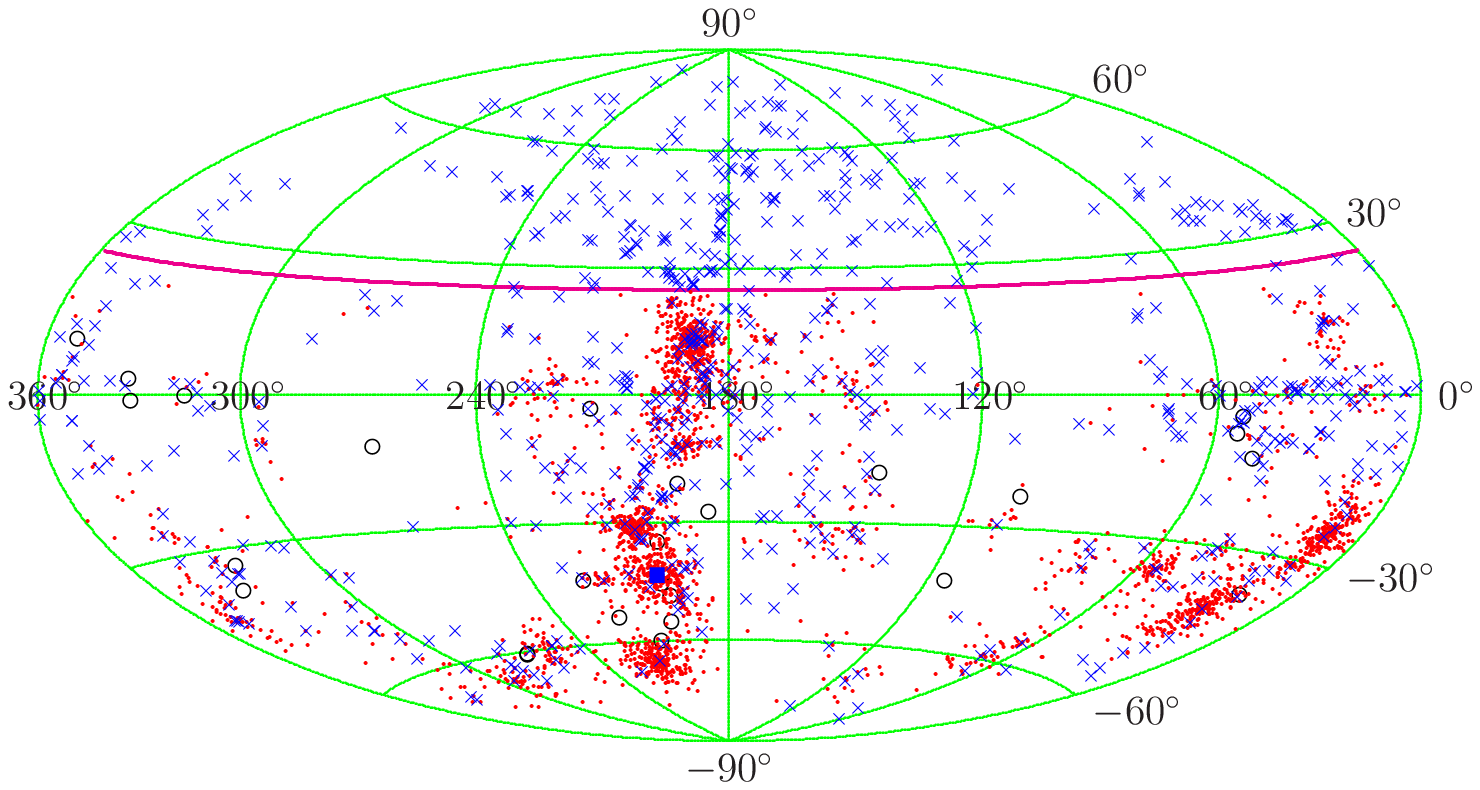}}
\par\vspace{5mm}\par
\centerline{\includegraphics[width=100mm]{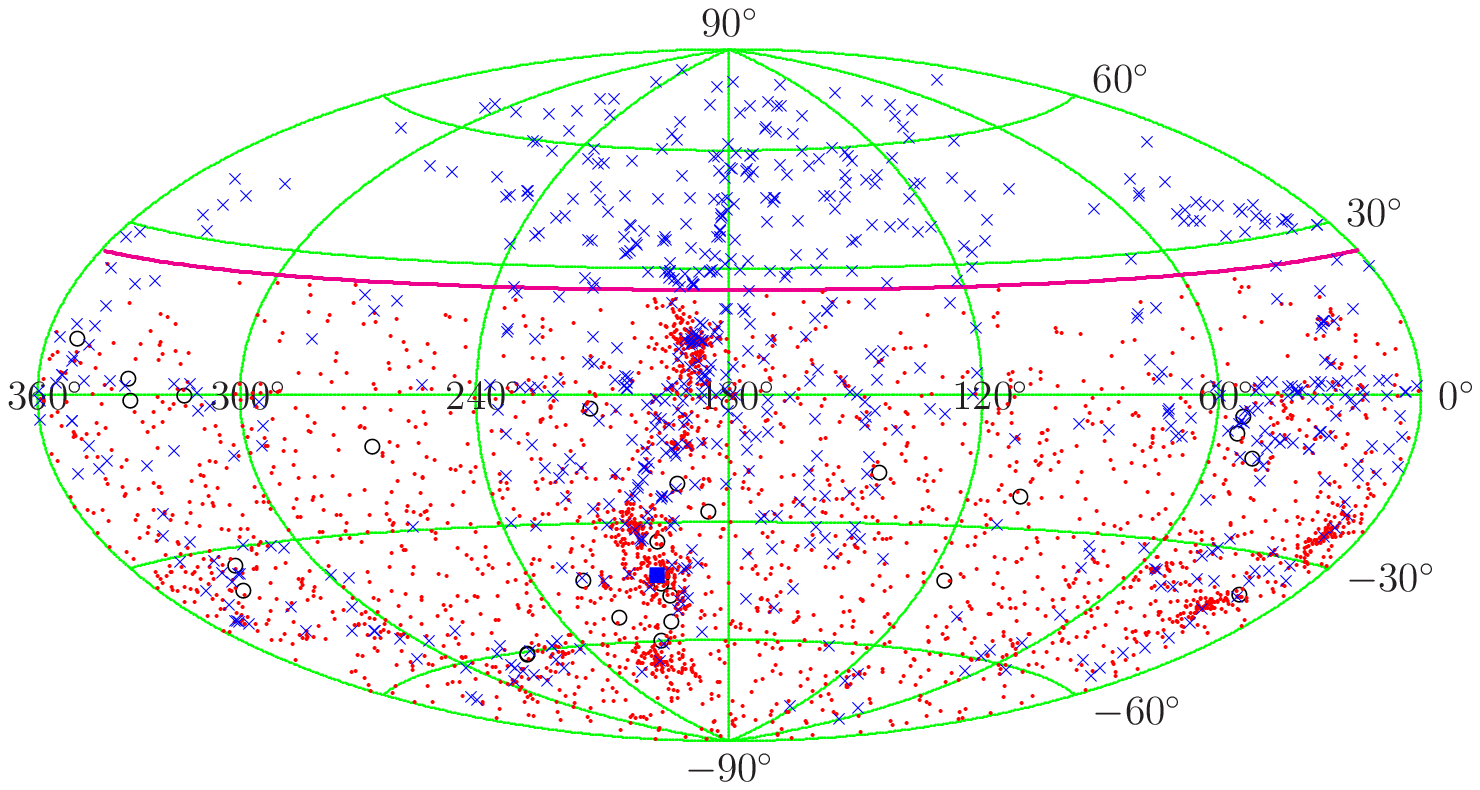}}
\caption{The distributions of simulated UHECR arrival directions
(2700 events, represented by red dots ({\color{red}.}))
obtained from the simple AGN model
with $f=1$ (upper panel) and $f=0.45$ (lower panel) in PAO case.
Mapping and other symbols are same as in FIG.~\protect\ref{skymap-all}.}
\label{skymap-sim}
\end{figure}

In the simple AGN model considered here, we assume that the fraction $f$ of
observed UHECR with $E\ge E_c$ are originated from the AGN within the distance $d_c$
and the remaining fraction $1-f$ from isotropic background contributions.
We call $f$ the AGN fraction and treat it as a free parameter of the model,
while its fiducial value is taken to be $f=f_0$.
In FIG.~\ref{skymap-sim}, we show the distributions of the UHECR arrival directions
expected from the simple AGN model with $f=1$ and $f=0.45$ as observed by PAO.

\section{The Correlation of AGN and UHECR data}
\label{sec4}

To illustrate our statistical methods,
we take the set of selected AGN as our reference set
and analyze the correlation between AGN and the arrival directions of UHECR
by using the CADD method.

\begin{figure}
\centerline{\includegraphics[width=100mm]{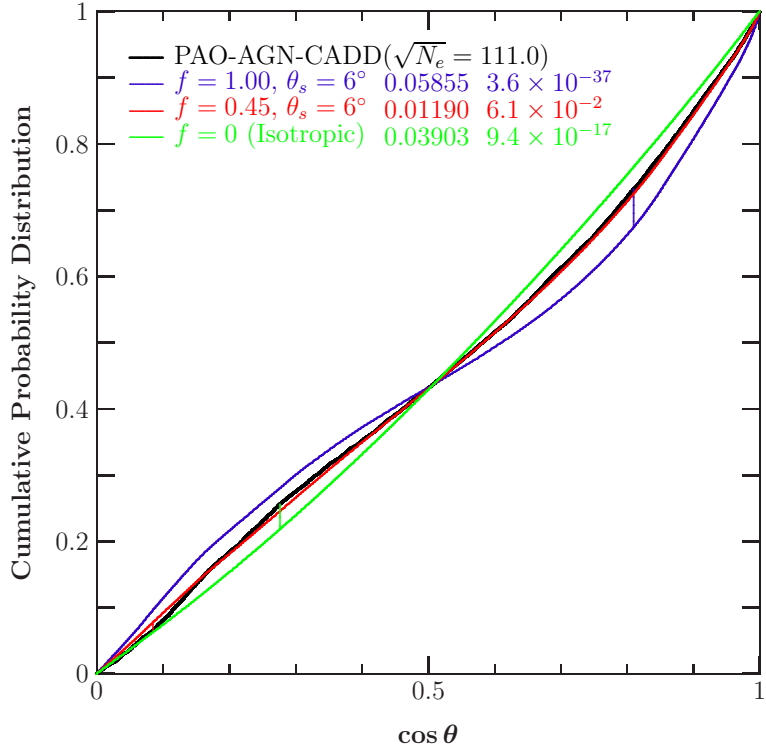}}
\caption{The cumulative probability distributions of CADD
using the AGN reference set.
The black line is for the observed PAO data,
the green line for the isotropic distribution, and
the blue and red lines for the simple AGN model
with a smearing angle $\theta_s=6^\circ$ and
a source fraction $f=1$ and $f=0.45$, respectively.
The vertical bars represent the KS statistics which are the maximum differences
between two CPDs.
The numbers in two right columns in the legend are the KS statistics and
the probabilities that the observed distribution of PAO arrival directions is obtained from the simple AGN models.}
\label{cpd-PAO}
\end{figure}

In FIG.~\ref{cpd-PAO} we compare the cumulative probability distribution (CPD)
of CADD between the selected AGN (with $d\le100\;{\rm Mpc}$) and
the arrival directions of UHECR from PAO data (with $E\ge5.7\times10^{19}\;{\rm eV}$)
with those expected from the simple AGN model
which is obtained through Monte-Carlo simulations.
Because the KS test is reparametrization invariant, we use the variable
$x=\cos\theta$ instead of the angular distance $\theta$ in the horizontal axis.
The CPD for the PAO data is the black line.
The green line represents the CPD for the isotropic distribution
(corresponding to $f=0$) of UHECR
and the blue and red lines represent the CPDs for the simple AGN model
with a smearing angle of $6^\circ$ for both
and the source fraction $f=1$ and $f=0.45$, respectively.
For comparing two different cumulative distribution functions $S_{N_1}(x)$
and $S_{N_2}(x)$, the KS statistic is \cite{NR}
\begin{equation}
D=\max\left|S_{N_1}(x)-S_{N_2}(x)\right|,
\end{equation}
thus it can be directly read off from the CPDs.
{}From the KS statistic $D$ of CADD, the probability that the PAO data are yielded from the given model is given approximately by
\begin{equation}
\hbox{Probability}=Q_{\rm KS}
\left(\left[\sqrt{N_e}+0.12+0.11/\sqrt{N_e}\right]D\right),
\end{equation}
where $Q_{\rm KS}(\lambda)=2\sum_{j=1}^\infty(-1)^{j-1}e^{-2j^2\lambda^2}$
and $N_e$ is the effective number of data points, $N_e=N_1N_2/(N_1+N_2)$.
When the number of observed UHECR events is $N_o$ and the number of AGN is $M$,
$N_1=N_oM$ for CADD, $N_1=N_o(N_o-1)/2$ for AADD, and $N_1=N_o$ for FEVD.
For $N_2$, we replace $N_o$ with $N_s$, the number of simulated UHECR events.
We compare the observed distribution, say $S_{N_1}$,
with the expected distribution $S_{N_2}$,
which is obtained from the model through simulations.
The expected distribution can be made accurate by increasing the number
of simulated UHECR events $N_s$.
We set $N_s=10^6$ for CADD and FEVD, and $N_s=10^4$ for AADD
for the reason of practical computation,
that is, it gives the accuracy sufficient for our purpose
in reasonable computation time,
though a little bit of fluctuations in calculating probabilities still remain.
Since $N_s$ is much larger than $N_o$, $N_e\approx N_1$.
The values of probability are shown in the upper-left corner in the figure.
We note that the CPD itself is also useful for comparing the correlations.
The steep rise of CPD near $\cos\theta=1$
means the strong correlation at small angles.
Comparing the CPDs, we see that the PAO data are much more correlated with the AGN
at small angles than the simple isotropic distribution.
The probability that the PAO data result from the isotropic distribution
is about $10^{-16}$.
But the PAO data are not correlated with AGN as much as required by
the simple AGN model with $f=1$ and a small smearing angle.
For example the probability that the PAO data are yielded wholly
from the AGN (that is $f=1$) with the smearing angle of $6^\circ$
is $2.1\times10^{-37}$, which is extremely small.

We can do the same kind of analysis now by using the FEVD or the AADD.
Different methods will give different probabilities for the same hypothesis,
showing the difference in their discriminating powers.
Thus, we combine three methods together to make better and reliable conclusions.

\subsection{Dependence on the AGN fraction and the smearing angle}

\begin{figure}
\centerline{
\includegraphics[width=80mm]{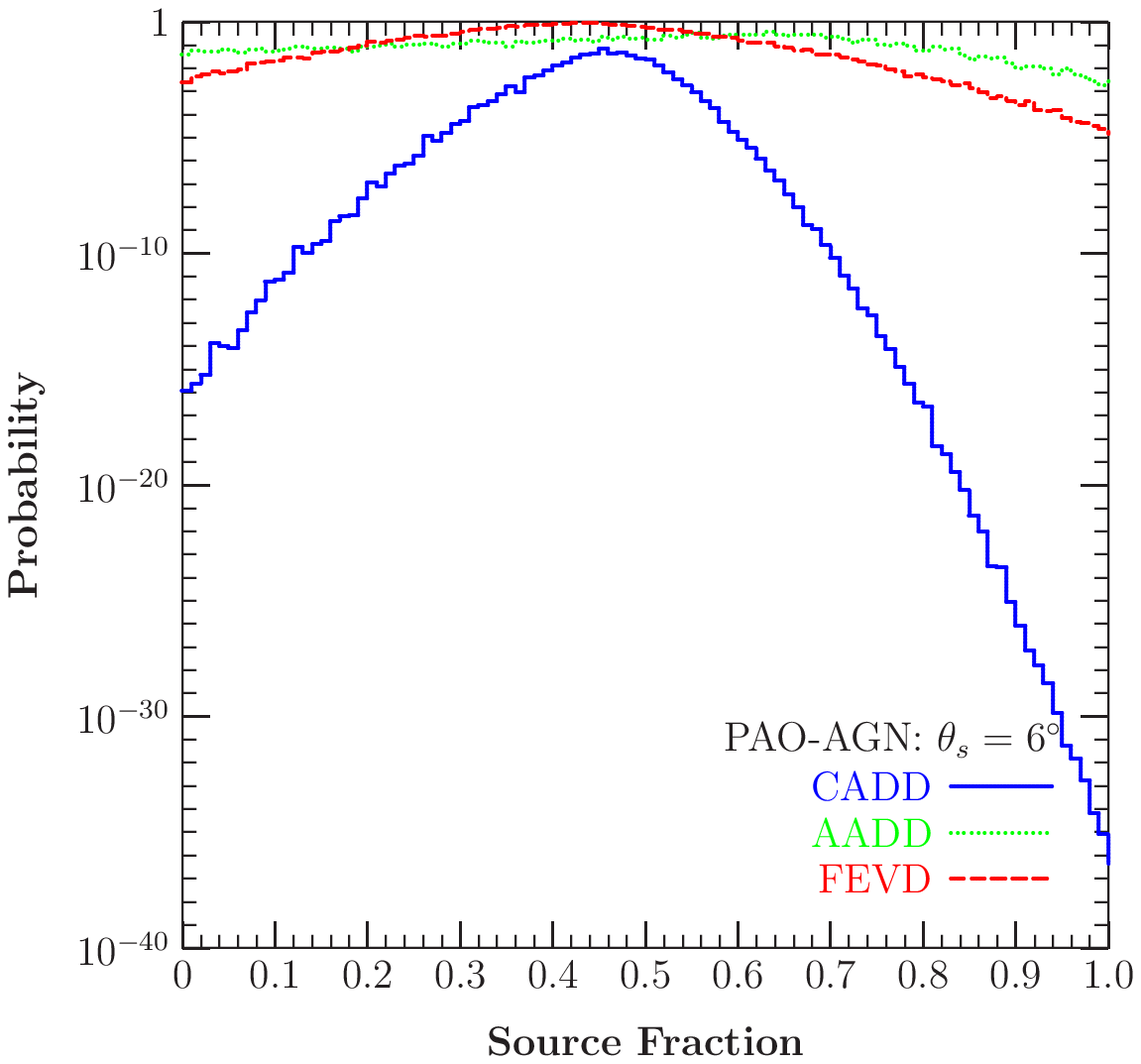}
\includegraphics[width=80mm]{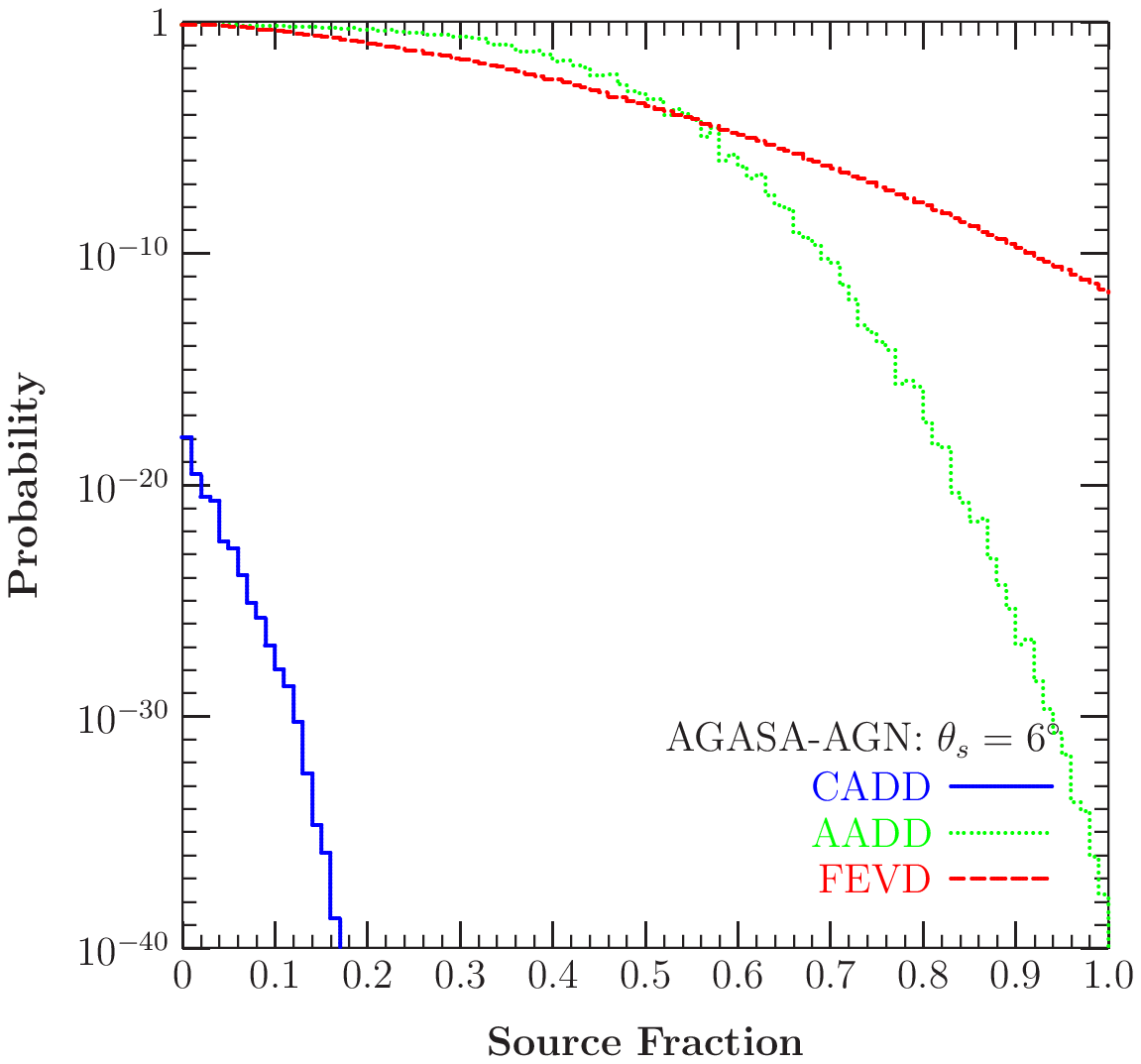}
}
\caption{Source fraction ($f$) dependence of probabilities
from three different analysis methods for the AGN and the PAO data.}
\label{prob-r}
\end{figure}

Now we examine the simple AGN model in detail.
Even though we introduced two parameters $f$ and $\theta_s$
for quite different motivations,
their effects on the distribution of UHECR arrival directions are not
clearly separated because both reducing $f$ and making $\theta_s$ larger
will make the distribution rather isotropic.
Let us first consider the source fraction dependence of the distribution.
In the simple AGN model,
while $f$ is the fraction of UHECR coming from the selected AGN,
the remaining $1-f$ is the fraction of UHECR coming from the isotropic background.
In FIG.~\ref{prob-r}, we showed the $f$ dependence of probabilities,
obtained by using CADD, FEVD, and AADD methods,
that the PAO data set and the AGASA data set are obtained
from the simple AGN model (with AGN within 100 Mpc) with the smearing angle $6^\circ$.
The CADD method gives the most stringent constraint
and it clearly excludes both limits of $f=0$ (complete isotropy)
and $f=1$ (all UHECR from selected AGN).
However,
a striking result is that for the PAO
data set the probability is peaked at $f\approx0.45$ with a value $0.064$.
In the previous section, we mentioned that
the energy cut of UHECR $E=5.7\times10^{19}\;{\rm eV}$ and
the distance cut of AGN $d_c=100\;{\rm Mpc}$
roughly determines the fraction of contribution from the inside
of the distance cut, $f\approx0.7$.
The peak is away from this value and
we need more isotropic components to get the best probability.
For the fixed value of $f\approx0.7$,
a good probability can be achieved by increasing the smearing angle,
as discussed below.

For the AGASA data set, we don't see any correlation with AGN.
Rather, the FEVD and AADD methods indicate that the observed AGASA data
are consistent with complete isotropy, even though the CADD method still gives
maximum but tiny probability.

\begin{figure}
\centerline{
\includegraphics[width=80mm]{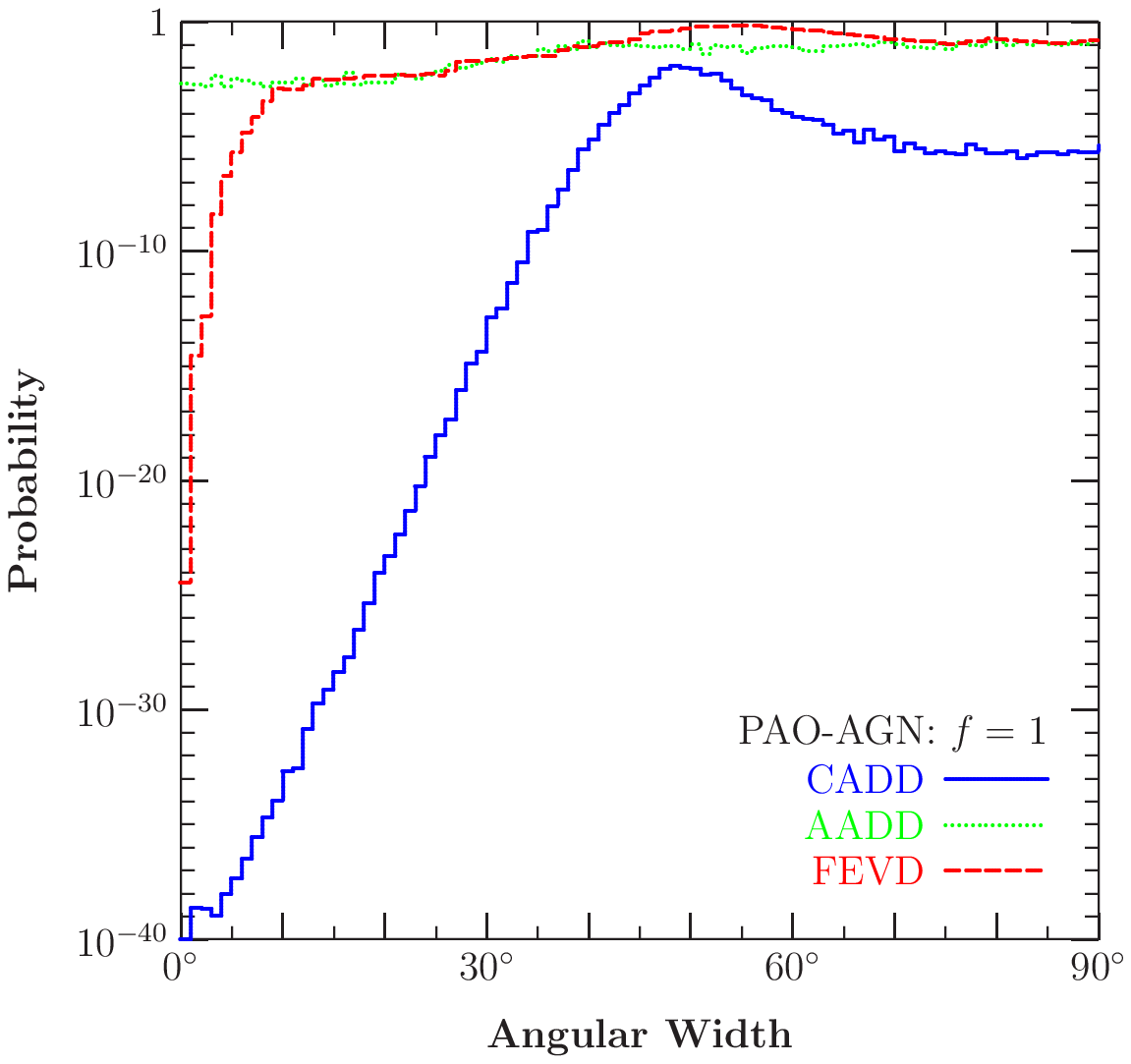}
\includegraphics[width=80mm]{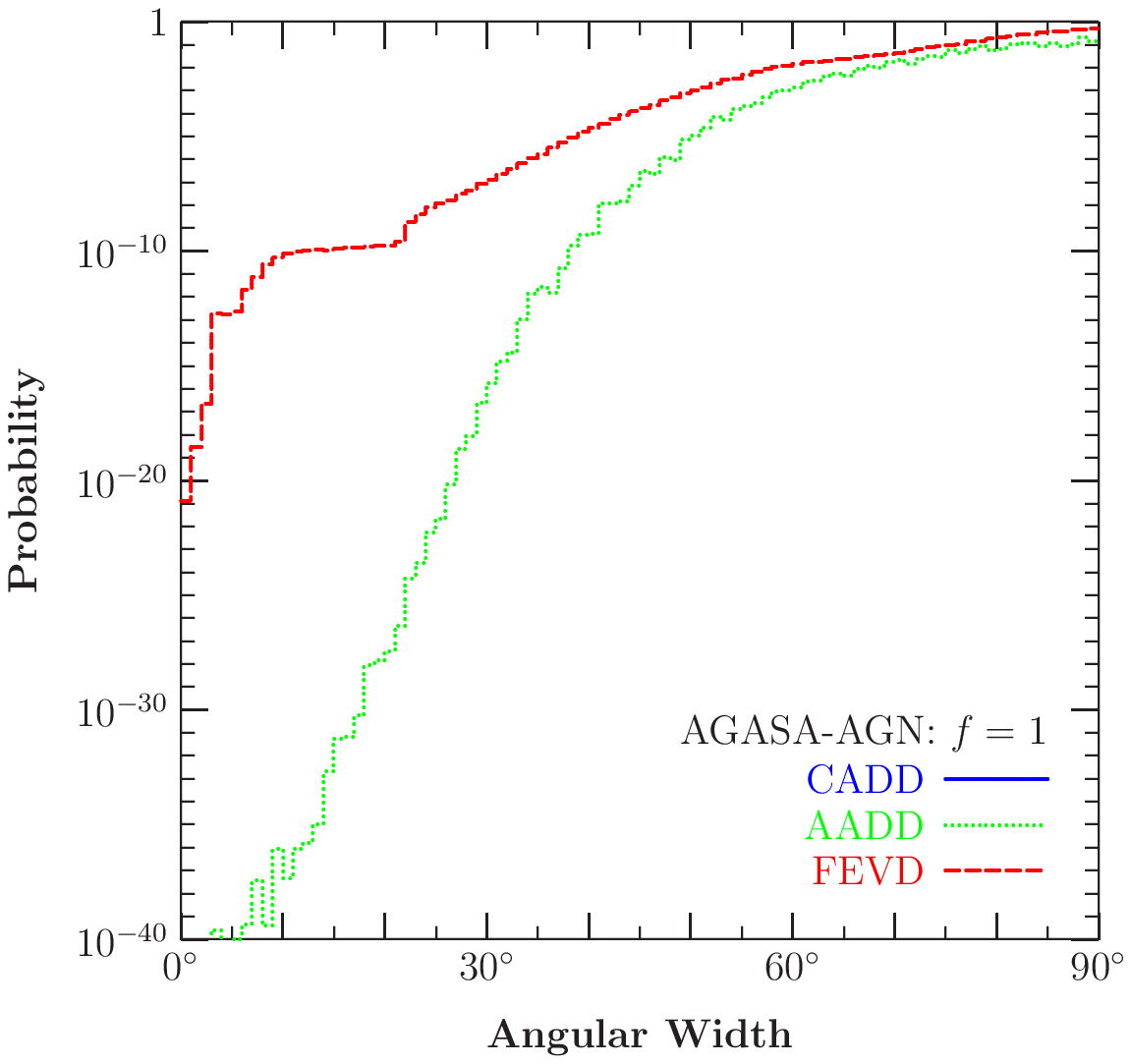}
}
\caption{Smearing-angle ($\theta_s$) dependence of probabilities
from three different analysis methods for the PAO data (left)
and the AGASA data (right) with the source fraction $f=1$.
In the AGASA result (right), the CADD line is not seen because it lies
below $10^{-40}$.}
\label{prob-w-r1}
\end{figure}

\begin{figure}
\centerline{
\includegraphics[width=80mm]{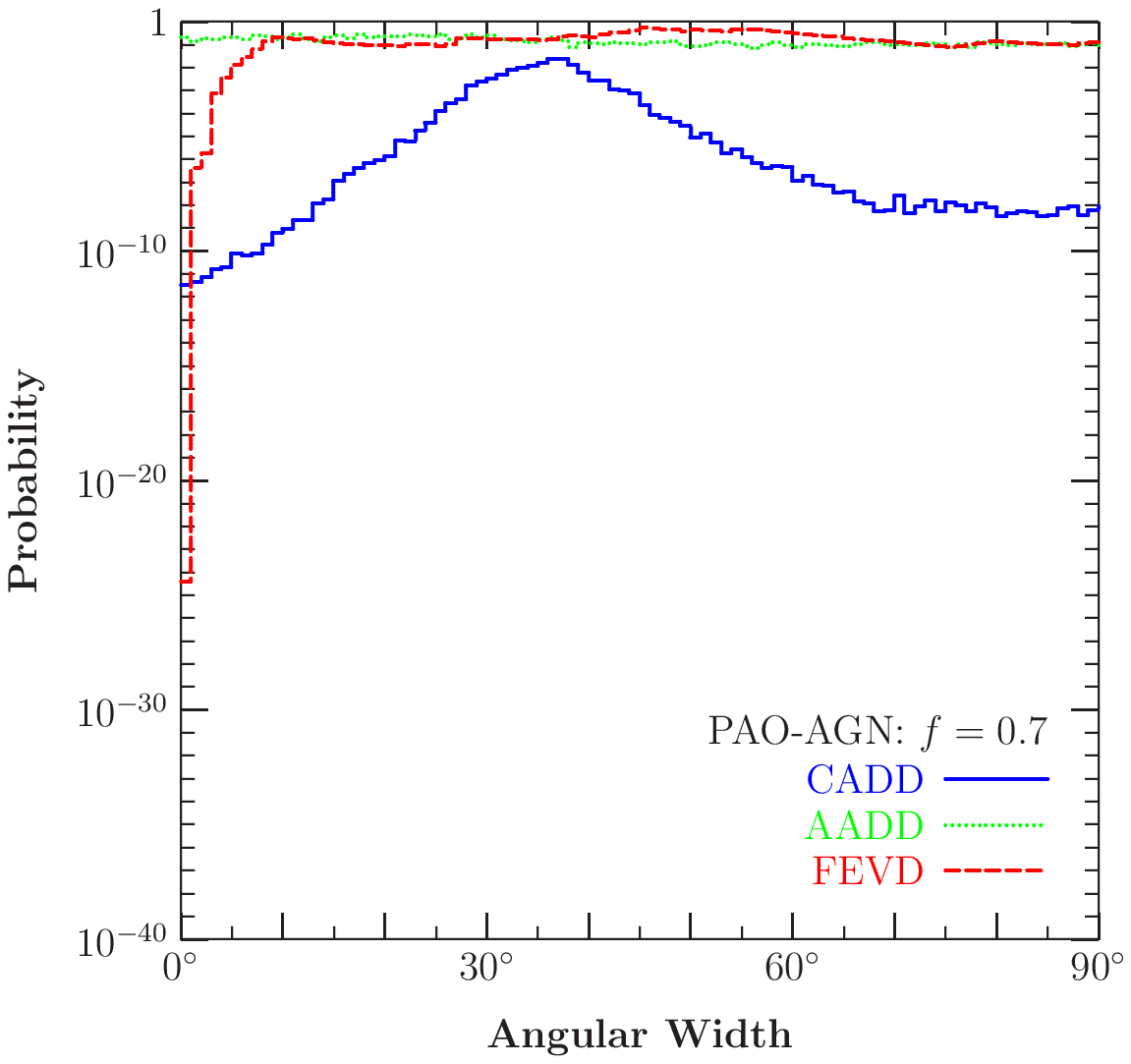}
\includegraphics[width=80mm]{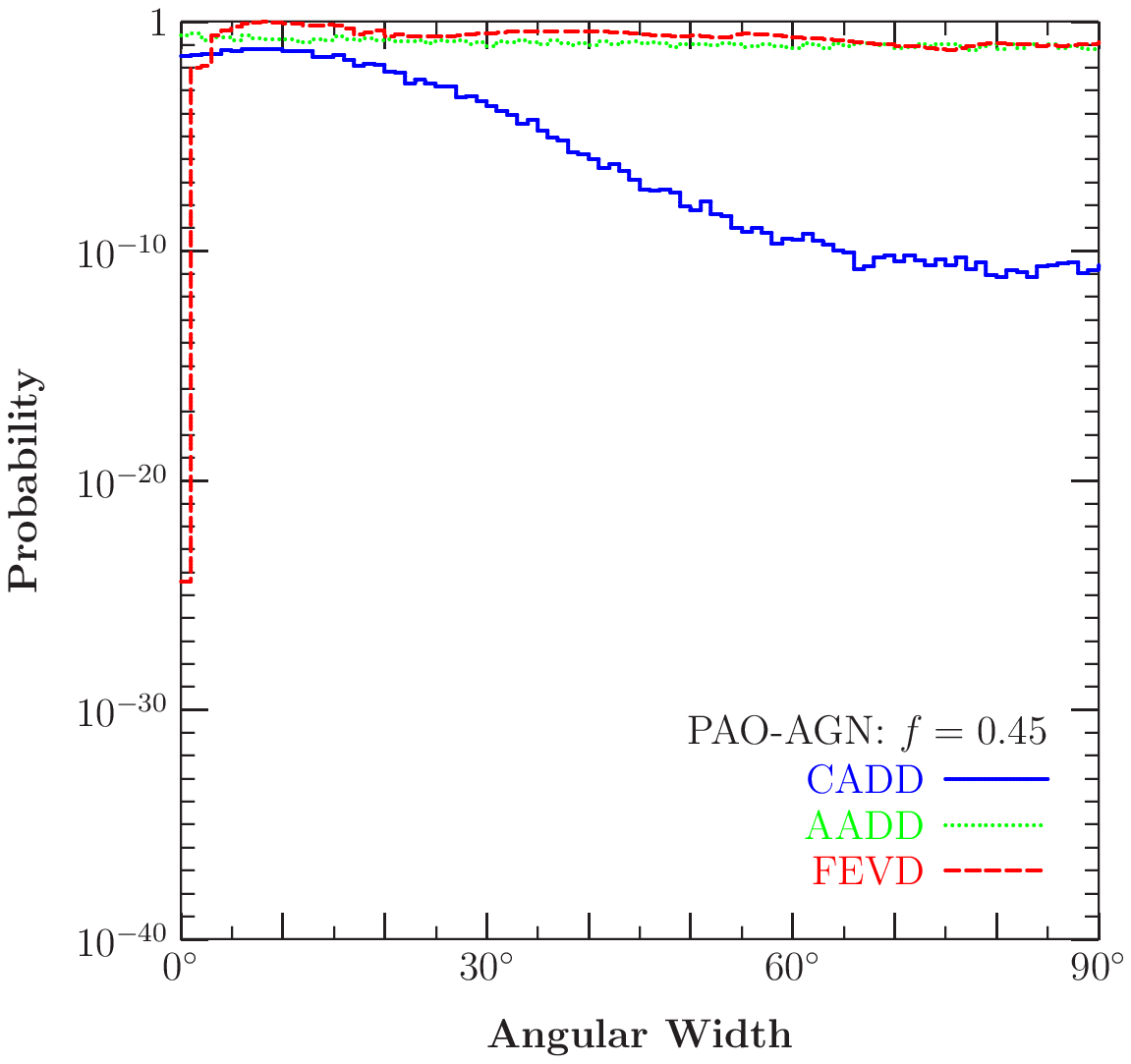}
}
\caption{Smearing-angle ($\theta_s$) dependence of probabilities
from three different analysis methods for the PAO data with the source
fraction $f=0.7$ (left) and $f=0.45$ (right).}
\label{prob-w-rx}
\end{figure}

The smearing angle $\theta_s$ in the simple AGN model is a free parameter,
though setting it a large value may needs some justification,
such as the existence of very strong intergalactic magnetic fields.
We also examine the smearing-angle dependence of the correlation analysis
and show the results in FIG.~\ref{prob-w-r1} and \ref{prob-w-rx}.
Here again the CADD method plays a main role,
while the FEVD method is sensitive at small smearing angles.
As the smearing angle increases, the distribution of the arrival directions
tends to be isotropic. In the above, we saw that the addition of isotropic
component to some extent improves the probability for the PAO data set.
Thus, we expect that increasing the smearing angle will also bring an improvement.
For $f=1$ (all UHECR from selected AGN), this happens for a rather large
value of the smearing angle and
the probability is peaked at $\theta_s=48^\circ$ with a value $0.016$
(See the left panel of FIG.~\ref{prob-w-r1}).
For $f=0.7$, which is the reference value for $E_c=5.7\times10^{19}\;{\rm eV}$
and $d_c=100\;{\rm Mpc}$, the peak value is 0.019 at $\theta_s=36^\circ$
(See the left panel of FIG.~\ref{prob-w-rx}).
If we set $f=0.45$, the small smearing angle is preferred
(See the right panel of FIG.~\ref{prob-w-rx}).

\subsection{Dependence on Distance-Bin}

In the above correlation analysis,
the number of observed UHECR is much smaller than the number of AGN.
Thus, only some subset of AGN can actually be responsible for observed UHECR.
Therefore, it is legitimate to classify AGN in some way
to narrow down the possible sources of UHECR among them.
Unfortunately, we don't have yet plausible criteria to classify AGN
into possible UHECR sources and not.
This is the reason why we included all AGN within a distance cut
assuming the equal UHECR luminosity for all AGN in the previous analysis.

\begin{figure}
\centerline{
\includegraphics[width=80mm]{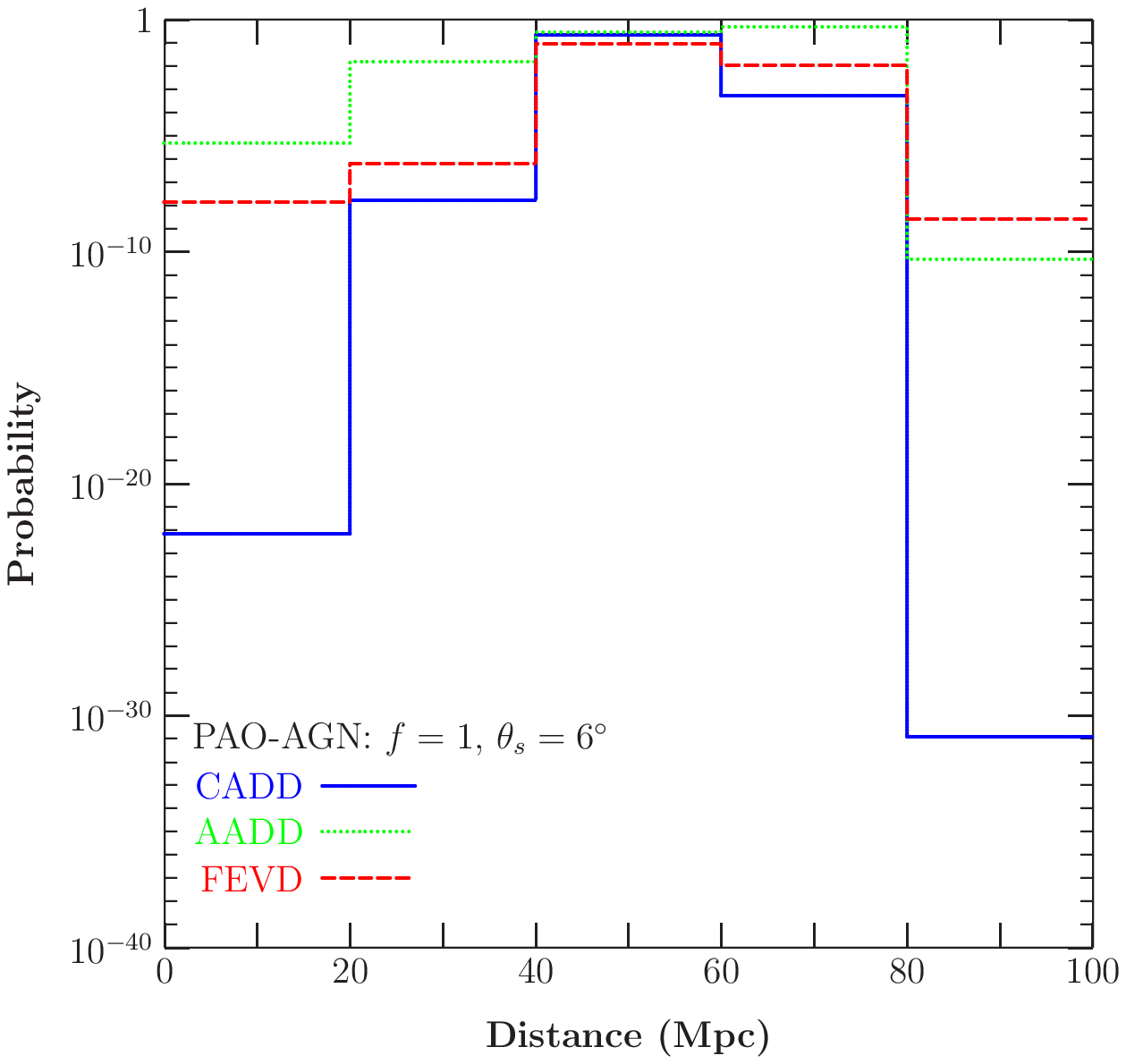}
\includegraphics[width=80mm]{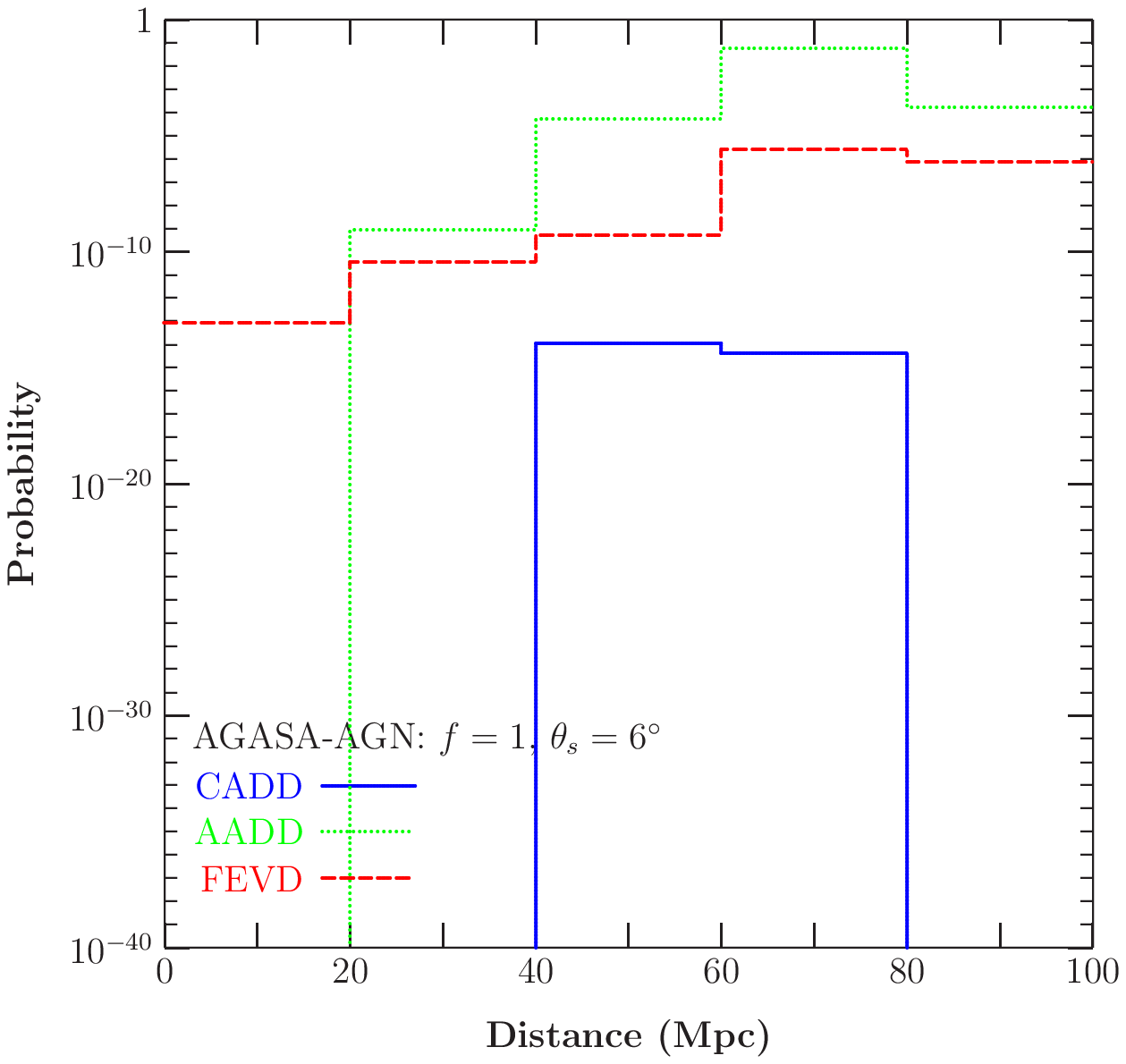}
}
\caption{Distance-bin dependence of probabilities for the PAO data (left panel)
and the AGASA data (right panels) with the distance bin size of 20 Mpc.}
\label{prob-d}
\end{figure}

\begin{figure}
\centerline{\includegraphics[width=100mm]{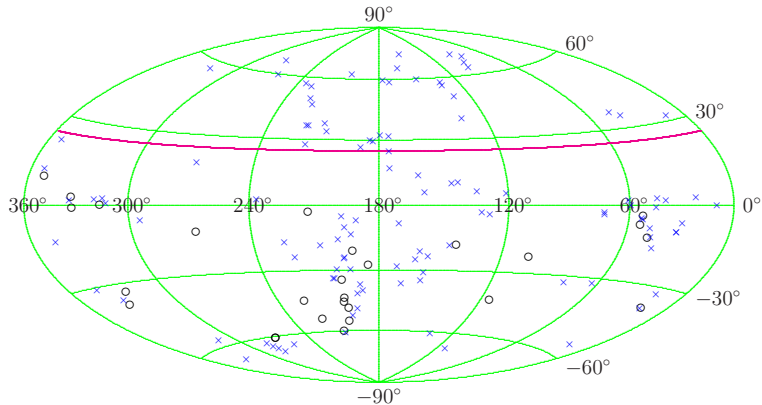}}
\caption{Skymap of AGN with $40\,{\rm Mpc}<d\le60\,{\rm Mpc}$ and
the arrival directions of UHECR with $E\ge5.7\times10^{19}\,{\rm eV}$
observed by PAO.}
\label{skymap-4060}
\end{figure}

Now we may consider a sophisticated hypothesis that
AGN with certain properties or within a certain region
are actually responsible for the UHECR.
As for the first and simple-minded step toward this direction,
we consider distance binning, though we don't have any reasonable motivation yet
except the exploratory analysis of the UHECR data.
FIG.~\ref{prob-d} shows the probabilities
by three analysis methods within the simple AGN model
with equally sized 20 Mpc distance-binning,
for which we set $f=1$, $\theta_s=6^\circ$
and use only those AGN within each bin as UHECR sources.
While we have tried several different bin sizes of 10, 20 Mpc etc.,
here we present the results for the 20 Mpc bin size
because it gives most interesting results.
For the AGASA data, nothing interesting happens.
But, quite interestingly, we find that with the PAO data
the bin with $40\,{\rm Mpc}<d\le60\,{\rm Mpc}$
gives remarkable probabilities for all three methods,
and this is manifestly seen with the bin size of 20 Mpc.
We may further try a single bin of arbitrary size to find the best
AGN set for the PAO data.
We find that the probability is maximized with the value $P_{\rm CADD}=0.27$
when we select AGN within 41 $-$ 63 Mpc bin.
For the comparison with FIG.~\ref{skymap-all},
FIG.~\ref{skymap-4060} shows the same skymap
as the upper panel in FIG.~\ref{skymap-all}
but for the AGN in the distance bin
$40\,{\rm Mpc}<d\le60\,{\rm Mpc}$.
We report this result just because it looks interesting,
though it seems difficult to make an appropriate interpretation yet.

\section{Discussion and Conclusion}
\label{sec5}

The previous analysis of the correlation between AGN and the arrival directions
of UHECR done by PAO group was based on the number of UHECR events
correlated with AGN, where `correlated' means that UHECR lie within
a certain angular distance $\psi$ from AGN \cite{Cronin:2007zz,Abraham:2007si}.
They used the data sets, AGN with $d\le71\;{\rm Mpc}$ and
UHECR with $E\ge5.7\times10^{19}\;{\rm eV}$, and the angular distance $\psi=3.2^\circ$
which minimize the probability of the null hypothesis that the distribution of
UHECR arrival directions is isotropic.
Under this condition, they found 20 out of 27 UHECR events are correlated
with AGN. The chance probability that this number occurs from the isotropic
distribution is $4.6\times10^{-9}$.
Instead of the correlated UHECR events,
our methods compares the reduced one-dimensional distributions,
CADD, AADD, and FEVD of the observed UHECR events and the expected ones
from the model, or the hypothesis.
When applied to the same data sets, our methods yield the chance probabilities
$P_{\rm CADD}\approx2.1\times10^{-9}$, $P_{\rm AADD}\approx4.9\times10^{-2}$, and
$P_{\rm FEVD}\approx2.5\times10^{-3}$ from the isotropic distribution.
These show that CADD is much better than FEVD and AADD
in revealing anisotropy of the PAO data.
The CADD method gives the result consistent with that of the PAO group.
The advantage of using CADD is that we can avoid the arbitrariness in
tuning the size of $\psi$ to minimize the chance probability.

Even though the null hypothesis of isotropic distribution can be ruled out
in these ways, it does not directly test the correlation between AGN and UHECR.
This fact was already pointed out in Ref.~\cite{Gorbunov:2007ja}.
For the direct test of correlation, we must set up an alternative hypothesis
that AGN are the sources of UHECR.
The null hypothesis we adopted is rather simple-minded in that
all AGN listed in the catalog are treated as equal sources of UHECR.
Our test reveals that the hypothesis that all UHECR
with $E\ge5.7\times10^{19}\;{\rm eV}$ observed by PAO
come from AGN with $d\le71\;{\rm Mpc}$
is also ruled out by the chance probability of $P_{\rm CADD}=1.3\times10^{-24}$.
It says that the observed UHECR are more correlated with AGN
than the simple isotropic distribution,
but not sufficiently correlated with AGN as required by that hypothesis.
Our result is consistent with that of Ref.~\cite{Gorbunov:2007ja}.

In conclusion, we developed the statistical methods for comparing two sets
of arrival directions of cosmic rays,
in which the two-dimensional distribution of arrival directions is reduced
to the one-dimensional distributions
such as CADD, AADD, and FEVD.
Then the probability that the observed arrival direction distribution
comes from the given hypothesis can be calculated by the standard
one-dimensional KS statistic.
We applied three methods together in combination
to the analysis of correlation between UHECR and AGN
to get improved and reliable conclusions.
We used UHECR with energies above $5.7\times10^{19}\;{\rm eV}$
observed by PAO and AGASA,
and AGN within the distance $100\;{\rm Mpc}$
listed in the 12th edition of VCV catalog.
As our analysis showed, CADD is more useful than FEVD and AADD
in testing the anisotropy and the correlation with AGN.

We set up the simple AGN model which assumes that all AGN within
a chosen distance cut are the equal sources of UHECR
having the same UHECR luminosity and smearing angle.
For the PAO data, our methods exclude both limits
that the observed UHECR are simply isotropically distributed
and that they are completely originated from the selected AGN.
However, we need to add the isotropic component to incorporate the contribution
from the outside of the distance cut and the smearing effect due to
the intergalactic magnetic fields.
Increasing isotropic component either through the background contribution
or through the large smearing effect improves the situation greatly
and makes the AGN hypothesis for UHECR sources a viable one.
But the required amount of the isotropic component is more than
the estimated contribution from the outside of the distance cut.
Thus, the rather large smearing angle is needed.
We also point out that restricting AGN within the distance bin of
$40\!-\!60\;{\rm Mpc}$ happens to yield a good probability
without appreciable isotropic component and large smearing effect.

For the AGASA data, we don't find any significant correlation with AGN.
Thus, as for the AGN hypothesis for the UHECR sources,
we have quite different and even conflicting conclusions
from PAO and AGASA experiments which mainly covers
the southern sky and the northern sky, respectively.
Of course, we cannot completely exclude the possibility that
the observed correlation with AGN for the PAO data is just a mere chance.
Now, the northern sky is being covered by Telescope Array (TA) experiment.
We hope the UHECR data from TA will resolve this problem.

[Note Added] As this work is completed, PAO released the new data of UHECR
\cite{:2010zzj}. Also new catalogs of AGN are available now
\cite{Abdo:2010ge,VCV:13}.
The results for these new data sets will be presented in a separate paper
\cite{KK}.

\section*{ACKNOWLEDGMENT}

This research was supported by Basic Science Research Program
through the National Research Foundation (NRF)
funded by the Ministry of Education, Science and Technology
(2009-0083563, 2010-0016307).

\end{document}